\documentclass[12pt]{iopart}

\usepackage{multirow}
\usepackage{graphicx}
\usepackage[acronym]{glossaries}
\usepackage{siunitx}
\usepackage{amsmath}
\usepackage[version=3]{mhchem}
\usepackage{harvard}
\citationmode{abbr}

\begin{document}
\bibliographystyle{dcu}

% --- ACRONYMS
\newacronym{SNR}{SNR}{signal-to-noise ratio}
\newacronym{TEPC}{TEPC}{tissue equivalent proportional counter}
\newacronym{MCA}{MCA}{multi channel analyzer}
\newacronym{LET}{LET}{linear energy transfer}
\newacronym{RBE}{RBE}{radio biological effectiveness}
\newacronym{FWHM}{FWHM}{full width at half maximum}

% --- AFFILIATIONS
\newcommand{\ati}{TU Wien, Atominstitut, Stadionallee 2, 1020 Wien, Austria}

\newcommand{\muw}{Medical University of Vienna, Department of Radiation Oncology, Spitalgasse 23, 1090 Wien, Austria}

\newcommand{\landsteiner}{Karl Landsteiner University of Health Sciences, Department of Oncology, Dr.-Karl-Dorrek-Straße 30, 3500 Krems an der Donau, Austria}

\newcommand{\hephy}{Institute of High Energy Physics of the Austrian Academy of Sciences, Nikolsdorfer Gasse 18, 1050 Wien, Austria}

\newcommand{\maus}{EBG MedAustron GmbH, Marie-Curie Straße 5, 2700 Wiener Neustadt, Austria}

%%%%%%%%%%%%%%%%% START %%%%%%%%%%%%%%%%%%

\title[Knopf et al.: Exploring offline pileup correction]{Exploring offline pileup correction to improve the accuracy of microdosimetric characterization in clinical ion beams}

\author{Matthias Knopf$^{1,*}$, Sandra Barna $^{2,3}$, Daniel Radmanovac$^{4}$, Thomas Bergauer$^{4}$, Albert Hirtl$^{1}$, Giulio Magrin$^{5}$}

\address{$^{1}$ \ati}
\address{$^{2}$ \muw}
\address{$^{3}$ \landsteiner}
\address{$^{4}$ \hephy}
\address{$^{5}$ \maus}

\ead{matthias.knopf@tuwien.ac.at}

\begin{abstract}
\textit{Objective:}
Microdosimetry investigates the energy deposition of ionizing radiation at microscopic scales, beyond the assessment capabilities of macroscopic dosimetry. This contributes to an understanding of the biological response in radiobiology, radiation protection and radiotherapy. Microdosimetric pulse height spectra are usually measured using an ionization detector in pulsed readout mode. This incorporates a charge-sensitive amplifier followed by a shaping network. At high particle rates, the pileup of multiple pulses leads to distortions in the recorded spectra. Especially for gas-based detectors, this is a significant issue, that can be reduced by using solid-state detectors with smaller cross-sectional areas and faster readout speeds. At particle rates typical for ion therapy, however, such devices will also experience pileup. Mitigation techniques often focus on avoiding pileup altogether, while post-processing approaches are rarely investigated.
\textit{Approach:}
This work explores pileup effects in microdosimetric measurements and presents a stochastic resampling algorithm, allowing for offline simulation and correction of spectra. Initially it was developed for measuring neutron spectra with tissue equivalent proportional counters and is adapted for the use with solid-state microdosimeters in a clinical radiotherapy setting.
\textit{Main results:}
The algorithm was tested on data acquired with solid-state microdosimeters at the MedAustron ion therapy facility. The successful simulation and reduction of pileup counts is achieved by establishing a limited number of parameters for a given setup.
\textit{Significance:} The presented results illustrate the potential of offline correction methods in situations where a direct pileup-free measurement is currently not practicable.
\end{abstract}

\vspace{2pc}
\noindent{\it Keywords}: microdosimetry, microdosimetric spectrum, pulse pileup, pileup correction

\submitto{\PMB}

\maketitle

\section{Introduction}

Microdosimetry is concerned with the assessment and description of the energy deposition of ionizing radiation in microscopic sites. This is of particular interest in the context of ion therapy, given the complicated interplay of different energy loss mechanisms for high \gls{LET} radiation. Ion therapy uses charged particles, such as protons and carbon ions, to irradiate tumors with high precision. Due to the targeted energy deposition, adverse effects on healthy tissue can be reduced. This requires a detailed planning of the treatment in advance. Radiation quality, essentially describing the type and energy of particles crossing a site ~\cite{Rossi_1986_Radiation_Quality,Menzel_2019_Radiation_Quality}, is currently considered in clinical treatment planning systems as the dose averaged LET, $LET_d$ \cite{Hagiwara_2020_LETd}. However, $LET_d$ is a calculated quantity and is susceptible to significant disparities ~\cite{Granville_2015_LET}. LET spectra are typically obtained from Monte Carlo simulations with differing outcomes depending on the simulated geometry, the scoring technique, the choice of physics models, different averaging techniques and the applied production cuts. The values between treatment centers show notable deviations due to the use of disparate methodologies and varying degrees of reporting ~\cite{Kalholm_2021_LET,Hahn_2022_LET}. Moreover, even a precisely known LET may be insufficient in predicting clinical outcomes. Recent studies indicate that different radiation qualities may result in a different \gls{RBE} despite featuring identical $LET_d$. Microdosimetry could serve as a more accurate predictor of radiobiological outcomes by incorporating the whole energy deposition spectrum into radiobiological modeling ~\cite{Guan_2024_uDos,Magrin_2023_State_of_the_Art}.\\

Microdosimetric data is typically obtained in the form of energy deposition spectra using a spectroscopic readout chain for pulse height analysis ~\cite{Parisi_2022_uDos_Review,ICRU_98_Microdosimetry}. Following a particle crossing in a proportional counter or solid-state detector, the signal charge is integrated using a charge-sensitive preamplifier. It is further shaped by a feedback network to improve the \gls{SNR}, and consequently digitized using a \gls{MCA}. This allows for measuring the energy imparted, $\Delta E$, per event. Due to the wide dynamic range of pulse heights in some microdosimetric measurements, multiple amplifiers with varying gain are typically used. In this work, however, all measurements were acquired using a single amplifier. The measured value for the energy imparted is divided by the mean chord length of the sensitive volume crossed, to obtain the lineal energy $y=\Delta E / \bar{\ell}_\text{path}$. The mean chord length $\bar{\ell}_\text{path}$ is defined as the mean distance a particle travels when traversing the sensitive volume. The lineal energy $y$ serves as an indicator for radiation quality in a microscopic volume ~\cite{ICRU_98_Microdosimetry}.\\

The objective of microdosimetry is to quantify the pattern of energy deposition at the microscopic scale. This inherently involves measuring small signal charges (down to several \SI{}{\femto\coulomb} per event), necessitating the use of low noise electronics to ensure sufficient spectral resolution. On the other hand, low noise electronics typically require shaping times on the order of microseconds \cite{Bertuccio_2023_Electronic_Noise} and thus are susceptible for pileup at high particle rates, impeding an accurate energy measurement. In the field of microdosimetry, gas-based \glspl{TEPC} have historically been the primary instrument of choice ~\cite{ICRU_98_Microdosimetry}. Solid-state devices offer faster readout speeds and smaller cross-sectional areas \SIrange[range-units = brackets, range-phrase = --]{0.002}{0.05}{\milli\meter\squared}, therefore reducing pileup ~\cite{Rosenfeld_2002_uDos_Hadrontherapy}. However, in clinical ion beams with particle rates on the order of \SI{e10}{\per\second} ~\cite{ICRU_93_Proton_Therapy}, even these small-scale microdosimeters will experience significant pileup.\\

As pileup does not have any physical significance in microdosimetric measurements, it should be reduced as much as possible to yield accurate single-particle energy deposition spectra. Under ideal conditions, a single particle entering the sensitive volume of the detector generates a charge cloud along its track, which is rapidly collected at the electrodes, forming a delta-like current pulse. It is amplified into a shaped pulse and digitized by the readout electronics before the next event occurs. Pileup can originate either during the charge collection in the detector (detector pileup) or as an artifact of the electronic processing (pulse pileup). In the following, only pulse pileup will be discussed, assuming the fast readout speeds achieved with thin solid-state detectors, which are on the order of \SI{1}{\nano\second}. This is significantly faster than the processed pulse shapes and the expected event time distribution on the small surface area. Pulse pileup thus refers to the partial or complete overlap of two or more shaped voltage pulses in the analog signal, which were generated by separate particles.

\begin{figure}[ht!]
\centering
\includegraphics[width=.8\textwidth]{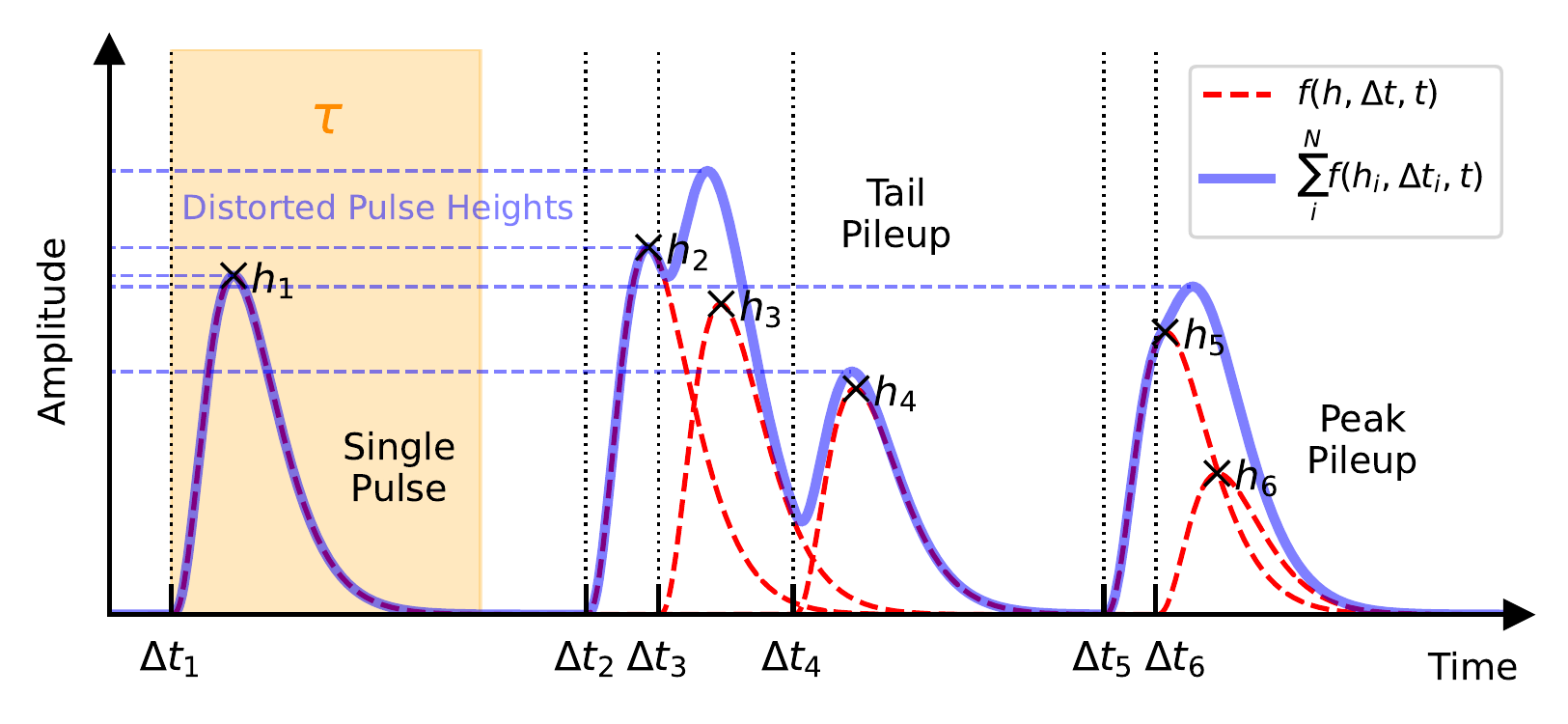}
\label{fig:pileup_formation}
\caption{Illustration of the formation of pulse pileup in the electronic readout chain. Individual pulses of heights $h_i$ are separated by inter-arrival times $\Delta t_i$. A pileup-free count is defined as two pulses separated by more than the pileup resolution time $\tau$. Due to partial overlap of pulses, the spectra are skewed and broadened (tail pileup), while the complete overlap (peak pileup) leads to the formation of double sum peaks. The dashed blue lines indicate the measured pulse heights from the sum of the individual waveforms.}
\end{figure}

Pulse pileup is sketched in figure \ref{fig:pileup_formation}. Peak pileup refers to the complete overlap of two or more events into a single registered pulse. In a spectrum containing distinct lines, peak pileup is most evident by the formation of double sum peaks. In extreme cases, even higher-order random sum peaks can be observed ~\cite{Blatt_1975_Sum_Peaks}. Tail pileup refers to the overlap of pulses with inter-arrival times higher than the rise time of a single shaped pulse. Individual events can still be resolved, but depending on their sequence and type of shaping this manifests as a broadening and skewing of the spectrum. In practice, a combination of both effects is observed, resulting not only in distortions of the spectrum, but also the potential appearance of artifacts.\\

Pileup correction is a complex undertaking, that, despite the best efforts, will never be able to perfectly remove pileup events from the final spectra. It is thus recommended that efforts are made to prevent pileup online during the measurement, whenever possible. A trivial solution is to lower the particle rate arriving at the detector by either lowering the detector cross section or modifying the dose rate. Modifying the particle flux from the accelerator is not always feasible, especially when the objective is to characterize clinical ion beams. In this context, it is crucial that the measurements accurately reflect clinical dose rates of up to $10^{10}$ \SI{}{\per\second}. Variations in extracted particle rates can affect the beam optics, altering the beam spot size and position. Moreover, monitoring systems in medical accelerators are designed to function reliably only at sufficiently high fluxes, when enough signal is being generated. This further emphasizes the need to maintain clinical conditions during measurements. At present, most microdosimetric setups are not equipped to measure pileup-free spectra at such particle fluxes.\\

A number of approaches have been put forth in various scientific disciplines to circumvent or correct pulse pileup in spectroscopic measurements ~\cite{Mohammadian_2020_Review_Pileup_Correction}.  Many spectroscopy systems employ a pileup-rejection scheme based on splitting the signal onto two channels after the preamplifier. A fast channel optimized for timing measurements identifies pulse arrival times and the slow channel prepares the signal for high resolution pulse height measurement with correspondingly long integration times. A delay line in the slow channel enables the fast signals to gate the acquisition in the event of pileup ~\cite{Bertolini_1964_Pileup_Circuit}. Such a discrimination circuit can be further enhanced by incorporating digital pulse processing ~\cite{Redus_2006_AMPTEK} or discriminating by pulse shapes ~\cite{Jordanov_2018_Shape_Discrimination,Glenn_2021_Shape_Discrimination}. Nevertheless, at exceedingly high rates, these methods have proven inadequate for the effective capture of pileup events. A promising approach in advanced pileup rejection can be based on an accurate recovery of arrival times, achieved by signal deconvolution ~\cite{Gadomski_1992_Deconvolution,Saxena_2020_Deconvolution}. The reconstruction of undistorted pulse shapes from the digitized pulse train has further been the subject of extensive investigation ~\cite{Drndarevic_1989_Scinti_4,Komar_1993_Scinti_5,Scott_2005_Scinti_1,Belli_2008_Scinti_3,Trigano_2015_general_algo,Scoullar_2011_general_algo}, also including machine learning methods ~\cite{Jeon_2022_Deep_Learning,Kim_2023_Deep_Learning}. However, it should be noted that some of these algorithms are unable to consider peak pileup or require specific pulse shapes not commonly used in commercial shaping amplifiers. Further these methods require a high \gls{SNR} that cannot be easily achieved in very thin solid-state detectors. Additionally, advanced pileup correction and rejection algorithms are not commercially available, nor are they straightforward to implement. Gas detectors do not benefit from these algorithms due to detector pileup resulting from their larger cross-sectional areas and the low drift velocities of positive ions.\\

One initial effort to assess the possibility of offline correction was made by Langen et al. ~\cite{Langen}. They proposed a stochastic algorithm, utilizing Poisson statistics, to resample the measured counts and subsequently correcting microdosimetric spectra. Since it showed promising results for correcting pileup in neutron data measured with a \gls{TEPC}, in this work we aim to explore its potential for solid-state microdosimeters in ion beams, using a diamond microdosimeter ~\cite{Verona_2018_Diamond}. The proposed algorithm potentially offers a quick and straightforward approach to model pulse pileup and to correct pulse pileup occurring during analog processing. The method can be achieved using a conventional analog spectroscopic readout chain without digital signal processing.\\

\section{Materials \& Methods}

Assuming Poisson-distributed inter-arrival times, the probability of two or more events adding up to a single detected event can be derived for a pulsed readout system ~\cite{Knoll}. The statistics focus on the number of physical events occurring during the pileup resolution time $\tau$. This parameter represents the minimum required separation between two shaped pulses in the analog signal, not to be classified as pileup. In terms of pulse pileup, a spectroscopic readout chain is best modeled as a non-paralyzable system, meaning that the arrival of new particles within $\tau$ does not extend the pileup resolution time of the initial pulse. The probability of observing $(m+1)$ pulses within this interval $\tau$ at an effective average particle rate $\alpha$ is given by

\begin{equation}
\label{eq:pileup_prob}
    P(m) = \frac{\left ( \alpha \tau\right )^m \exp{(-\alpha \tau)}}{m!} \ \rightarrow p = P(m\geq 1) = 1-P(0).
\end{equation}

It follows, that the total pileup probability $p$, i.e. two or more events in a count $P(m\geq 1)$, can be given as a function of rate $\alpha$ and resolution time $\tau$. Together, they form the pileup magnitude $\alpha \tau$. Equation \eqref{eq:pileup_prob} allows for an estimate of the pulse pileup expected in a particle beam with Poisson distributed arrival times. A square slab detector centered in the beam encounters an average effective particle rate $R_\text{eff}$. For a Gaussian transversal distribution of particles, the effective particle rate can be calculated from the total particle rate $R$ and the ratio of the intensity arriving at the sensitive volume $I_\text{SV}$ to the total intensity over the whole beamspot $I_\text{Spot}$. The intensities are in turn calculated from the beam spot size $\sigma$ and the side length $L$ of the detector as

\begin{equation}
\label{eq:ion_beam_est}
    R_\text{eff} \approx R\cdot \frac{I_\text{SV}}{I_\text{Spot}} =
    R \cdot \frac{
    \int_{-L/2}^{L/2}\int_{-L/2}^{L/2}  \exp{\left (-\frac{x^2+y^2 }{2\sigma^2} \right )} \mathrm{d}x \mathrm{d}y
    }{
    \int_{-\infty}^\infty \int_{-\infty}^\infty \exp{\left ( -\frac{x^2+y^2}{2\sigma^2} \right )} \mathrm{d}x \mathrm{d}y
    }.
\end{equation}

\begin{figure}[htbp]
\centering
\includegraphics[width=.9\textwidth]{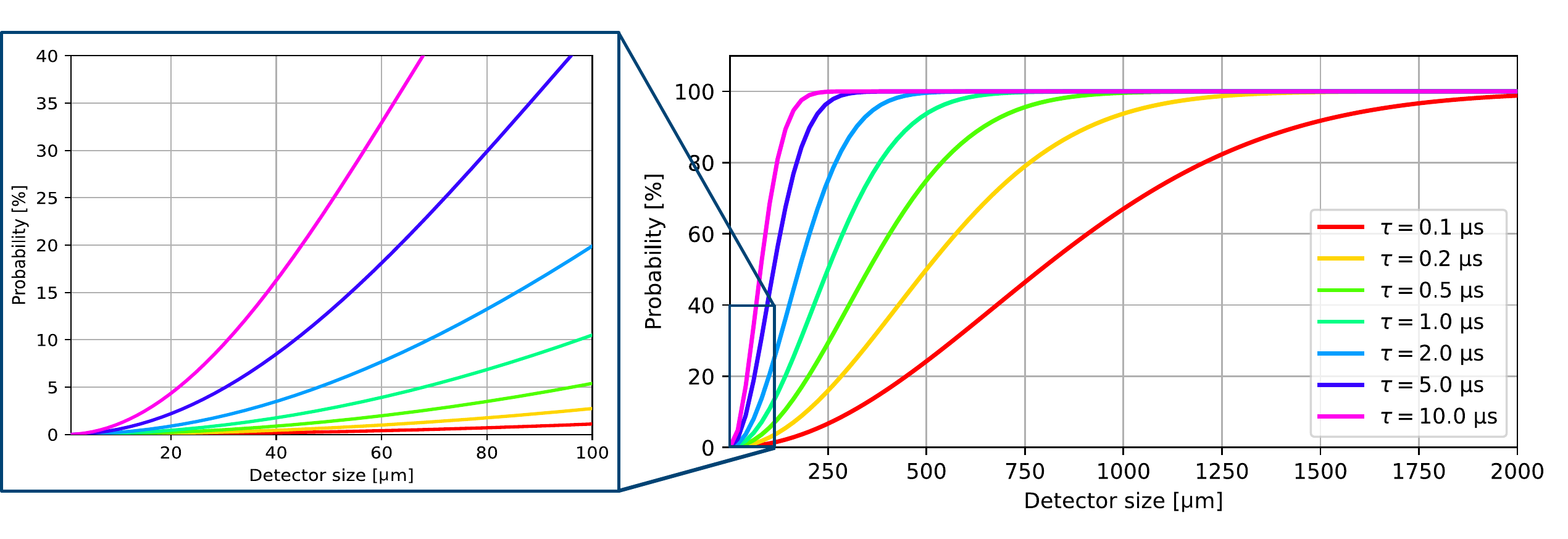}
\label{fig:poisson_prob}
\caption{Estimation of the pileup probability in a Poisson distributed particle beam on square detectors of different sizes. Pileup is defined as the arrival of two or more particles within the pileup resolution time $\tau$, leading to an overlap of the pulses. The plot illustrates the conditions for a \SI{5}{\milli\meter} \gls{FWHM} Gaussian beam spot with $10^8$ particles/s. The relevant range for state-of-the-art microdosimeters is shown in the zoomed plot to the left. At pileup resolution times $\Or(\SI{}{\micro\second})$, pileup contributions are notable at small cross sections. \label{fig:est_ion_beam}}
\end{figure}

\noindent Figure \ref{fig:est_ion_beam} shows the result of \eqref{eq:pileup_prob} and \eqref{eq:ion_beam_est} for a situation representative of the proton and carbon-ion beams at the MedAustron facility ~\cite{Grevillot_2020_MedAustron_Commissioning} with a particle rate of \SI{e8}{\per\second} on a Gaussian beamspot of \SI{5}{\milli\meter} \gls{FWHM} ~\cite{Ulrich_Pur_2021_Commissioning,Grevillot_2018_MAUS}. Even with small cross-sectional areas and short shaping times, pileup remains a concern for state-of-the-art microdosimeters. Due to intensity fluctuations and particle bunching, non-Poissonian ripples in the rate distribution are expected in a real beam, which leads to a further increase in the probability of pileup ~\cite{DeFranco_2021_MAUS_Intensity,Waid_2024,Torino_Rate_Meas}.\\

A resampling algorithm can be used to simulate the effect of pileup in microdosimetric spectra ~\cite{Langen}. For this, pulse height analysis is modeled in the time domain as a twofold random process. $N$ events with random pulse heights are $h_i$ separated by random time intervals $\Delta t_i$, as shown in figure \nolinebreak \ref{fig:pileup_formation}. Due to shaping, each event yields a similar pulse shape $f(h, \Delta t, t)$, differing only in amplitude $h$. The individual signals add up to a continuous pulse train $v(t) = \sum_i^N f(h_i, \Delta t_i, t)$ to be digitized. In a first step, the respective probabilities $P(m)$ for pileup of different order $m$ are calculated according to equation \eqref{eq:pileup_prob}, with the parameters $\alpha$ and $\tau$ known in advance. These probabilities are then multiplied by the total number of counts in the spectrum to determine the number of affected events. Unaffected counts are directly transferred to the output histogram, while affected pulse heights are randomly sampled from their frequency distribution. Depending on the order of pileup, tuples of $m$ events are grouped together and their individual offsets in time are sampled within $m \cdot \tau$ from the first pulse in the tuple. Next, the sum of the pulses per group is calculated by superposing the unit pulse shapes scaled with the corresponding sampled pulse heights. The unit pulse shape refers to the output of the shaping network normalized in height. A popular choice for pulse shaping are $CR-(RC)^n$ or "semi-Gaussian" shaping circuits, due to their relatively high \gls{SNR} and short tail pulses ~\cite{Radeka}. At equal time constants $T$, the semi-Gaussian waveforms can be described as

\begin{equation}
    V_\text{out}(t) = \frac{V_\text{max}}{n!} \left ( \frac{t}{T}\right )^n \exp{\left (-t/T \right )}.
\end{equation}

By adding $RC$ stages the pulse becomes more symmetrical until it converges into a Gaussian shape. For commercial analog shapers, the shaping circuit might not be a simple $RC$ circuit and the waveforms are difficult to describe analytically. As a result, their characteristics must be determined using an oscilloscope. In the end, peak heights are determined from the sum, allowing for the simulation of both peak and tail pileup with the possibility of peaks being completely buried. The distorted pulse heights are transferred to the output histogram.\\

As demonstrated by Langen et al \cite{Langen}, the simulation of pileup can be used to correct spectra:\\

\begin{enumerate}
    \item Using the algorithm described above, further pileup is simulated into the spectrum in question with a given $\alpha$ and $\tau$. $\tau$ is given by the unit pulse shape. $\alpha$ is either determined from prior analysis, estimated as the effective rate expected in this measurement or set to a small value ($\alpha \tau \ll 1$) for iterative correction.
    \item After pileup simulation using these parameters, the bin-wise difference between the measured input spectrum and the simulation output spectrum, the pileup vector, is calculated.
    \item The pileup vector is subsequently subtracted from the input spectrum, thereby providing an improved guess for the pileup-free version of the given spectrum.
\end{enumerate}

If the pileup parameters $\alpha$ and $\tau$ have been previously determined, sufficient correction can be achieved in one cycle (single step). Should this not be the case, an iterative correction process is started, repeating the steps above. If one has access to a pileup-free spectrum, the process is then repeated until convergence with this "ideal" spectrum. Convergence can be estimated using a least-squares difference between the respective input and output spectra. Once $\alpha \tau$ and the unit pulse shape have been established for a given setup, future spectra can be corrected using the same parameters.\\

The algorithm was tested using data taken at the MedAustron facility. Four separate sets of microdosimetric spectra were measured using the same acquisition chain, each set at a different radiation quality. Within each set, measurements were performed at two different particle rates. Each set of spectra was measured back to back using the same extraction method to ensure that differences in the shapes of a set of spectra can only be attributed to the particle rate. The detector's positioning at the center of the beamspot was verified using EBT3 films. The readout chain consisted of an Amptek CoolFET (A250CF) charge sensitive preamplifier with an ORTEC 671 spectroscopic shaping amplifier and an ORTEC 928 MCB multichannel analyzer (12 bit). The shaping amplifier features a pileup rejection (PUR) circuit employing fast-slow channel discrimination and a secondary rejection logic. The detector was a \qtyproduct{200 x 200 x 2}{\micro\meter\cubed} diamond microdosimeter ~\cite{Verona_2018_Diamond,Magrin_2020_Diamond_Measurements}. Reverse bias was supplied using a Rhode \& Schwarz HCM8043 bench-top power supply via a bias-tee integrated into the preamplifier. The unit pulse shape was determined as a Gaussian pulse with a shaping time $T=$\SI{1.7}{\micro\second} and a pileup resolution time $\tau =$\SI{7.5}{\micro\second} via least squares fitting of waveforms acquired using an oscilloscope, while signals were applied to the test input of the preamplifier. Microdosimetric spectra were obtained using \ce{^12C^6+} ions at 120 MeV/u and 238.6 MeV/u at both clinical rates, hereafter referred to as high flux, and using a degrader setting to limit the particle rate to 10\%, referred to as low flux. Internal measurements from MedAustron indicate a total particle rate of approximately $4.3\cdot10^7$ \SI{}{\per\second} in the high-flux setting. The particle rate for the low flux setting is thus approximately $4.3\cdot10^7$ \SI{}{\per\second}. Degrading is achieved by limiting the number of particles injected into the accelerator with a physical grid ~\cite{MAUS_Carbon_Comissioning}. A single beam spot of approximately \SI{6}{\milli\meter} \gls{FWHM} was used, and the detector was placed in the isocentre. Measurements were taken in the plateau region of the depth-dose curve, both without a phantom and using RW3 ("solid water") plates with a water-equivalent thickness of \SI{28.3}{\milli\meter} to position the detector at the Bragg peak for different radiation qualities. The position of the Bragg peak was previously established using the PeakFinder (PTW, Germany).\\

The algorithm operates directly on the digitized, linearly binned pulse heights. Details on the calculation of the microdosimetric representations of pulse height spectra can be found in ~\cite{Rossi_1996_Microdosimetry_and_its_Applications,ICRU_98_Microdosimetry}. The microdosimetric lineal energy representation

\begin{equation}
    yd(y) = \frac{y^2}{\bar{y}_F} f(y)
\end{equation}

is calculated after the pileup correction and energy calibration from the frequency lineal energy distribution $f(y)$, corresponding to the pulse height measurement. Further, the frequency average lineal energy $\bar{y}_F=\int_{y_0}^\infty yf(y) \mathrm{d}y$ and the dose average lineal energy $\bar{y}_D=1/\bar{y}_F\int_{y_0}^\infty y^2f(y) \mathrm{d}y$ are calculated above the noise threshold $y_0$. The microdosimetric representation $yd(y)$ uses logarithmic binning to improve the visibility of spectral differences.  All spectra were calibrated using carbon edge calibration, scaling the energy axis to the maximum energy that can be transferred by a single carbon ion in the given detector material, as described in ~\cite{Meouchi_2022_Uncertainties}. An additional spectrum was measured for this purpose at the distal edge of the Bragg curve, using RW3 plates with a water equivalent thickness of \SI{30}{\milli\meter}. This spectrum showed a distinct carbon edge, which was used to determine the scaling factor for the energy axis. A lower cutoff was set in all spectra to only include the main part of the spectrum and avoid sampling from the low energy tail formed by electronic noise and delta ray production outside of the sensitive volume ~\cite{Magrin_2022_Delta_Escape}. The spectra follow the Landau / Gaussian distribution typical for measurements of high energy ions. When sampling the affected events from the spectrum, a cutoff has to be set at the first empty channel to avoid sampling from empty energy bins not containing spectral information. All spectra measured in the plateau for 120 MeV/u ions were cut at $y_0=$ \SI{48}{\kilo\electronvolt\per\micro\meter}. The spectra measured at the Bragg peak for 120 MeV/u ions were cut at $y_0=$ \SI{99.6}{\kilo\electronvolt\per\micro\meter}. All spectra measured in the plateau for 238.6 MeV/u ions were cut at $y_0=$ \SI{31.8}{\kilo\electronvolt\per\micro\meter}.\\

\section{Results}
\begin{figure}[htbp]
\centering
\includegraphics[width=.375\textwidth]{fig3a_very_low_flux.pdf}
\qquad
\includegraphics[width=.5\textwidth]{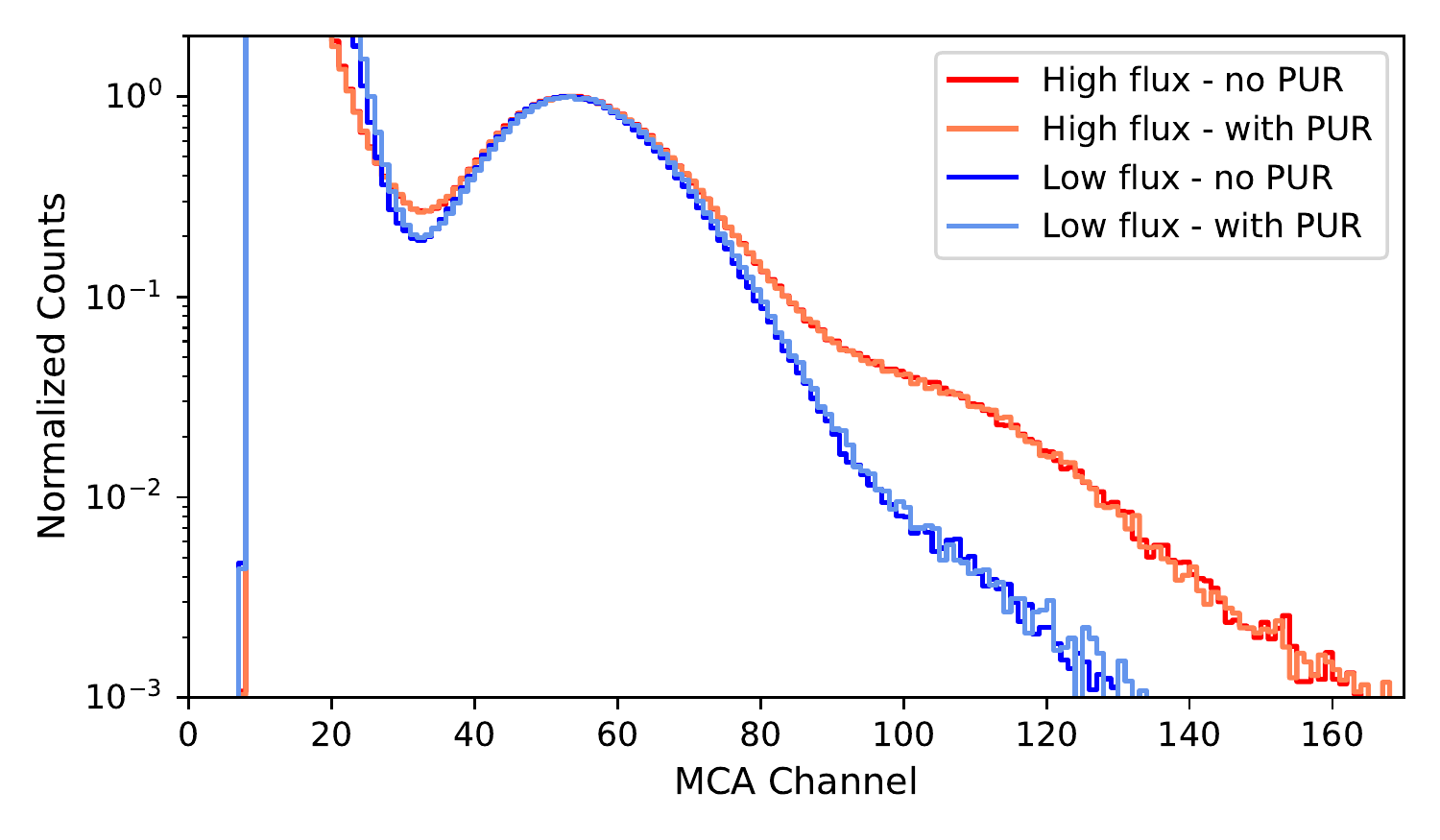}
\caption{A) Measured high flux, low flux spectra and extremely low flux spectra for \ce{^12C^6+} ions at 238.6 MeV/u. B) Measured high flux and low flux spectra for \ce{^12C^6+} ions at 120 MeV/u. The spectral differences due to pileup are evident by the formation of a double sum peak around channel 110. Both measurements with (light colors) and without the internal pileup rejection logic (rich colors) are shown. The pileup rejection circuit (PUR) was not able to reject pileup events at the rate encountered in the high flux measurement, as seen from the overlap of the spectra. \label{fig:linear_axis_spec}}
\end{figure}

\begin{figure}[htbp]
\centering
\includegraphics[width=.45\textwidth]{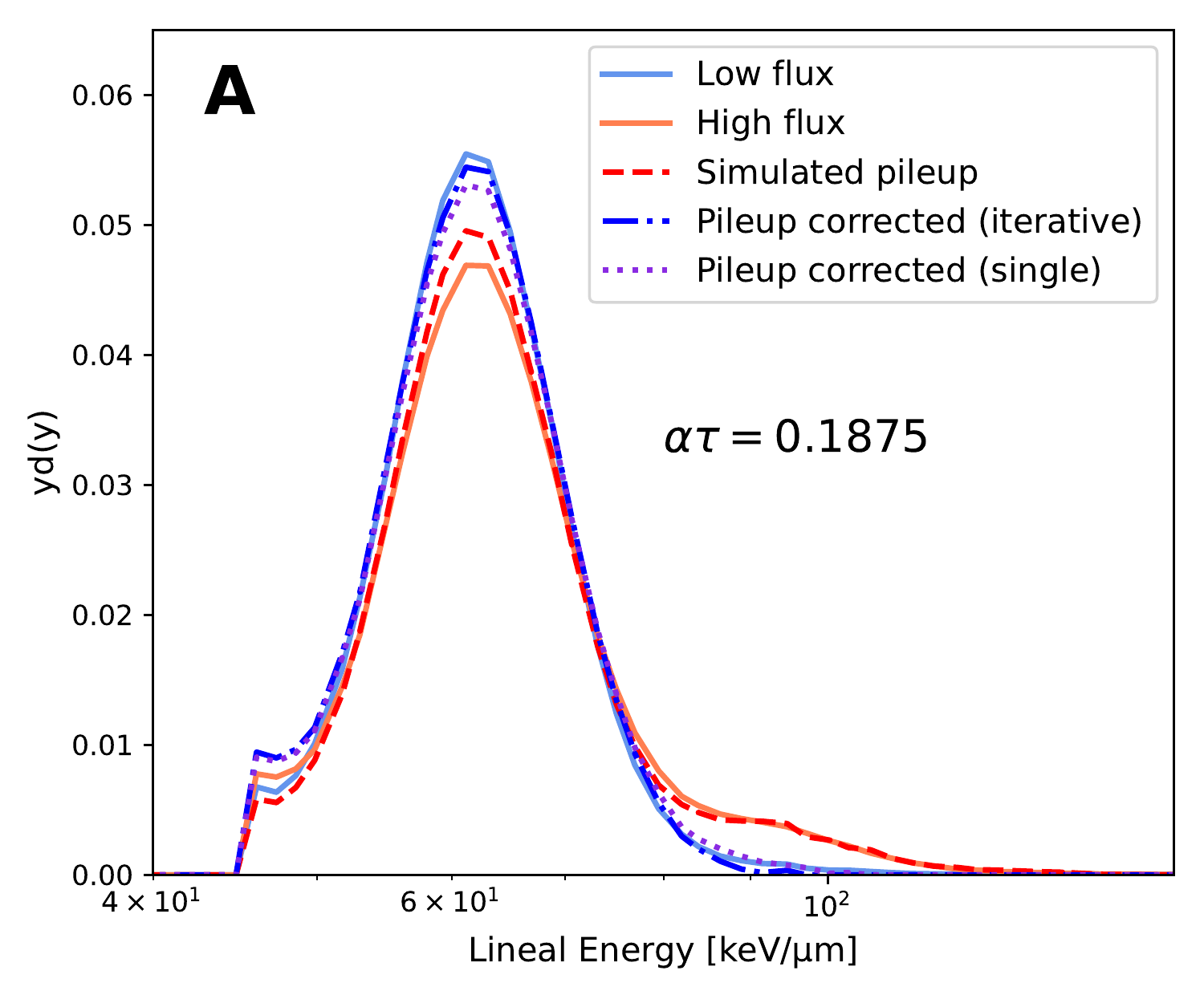}
\qquad
\includegraphics[width=.45\textwidth]{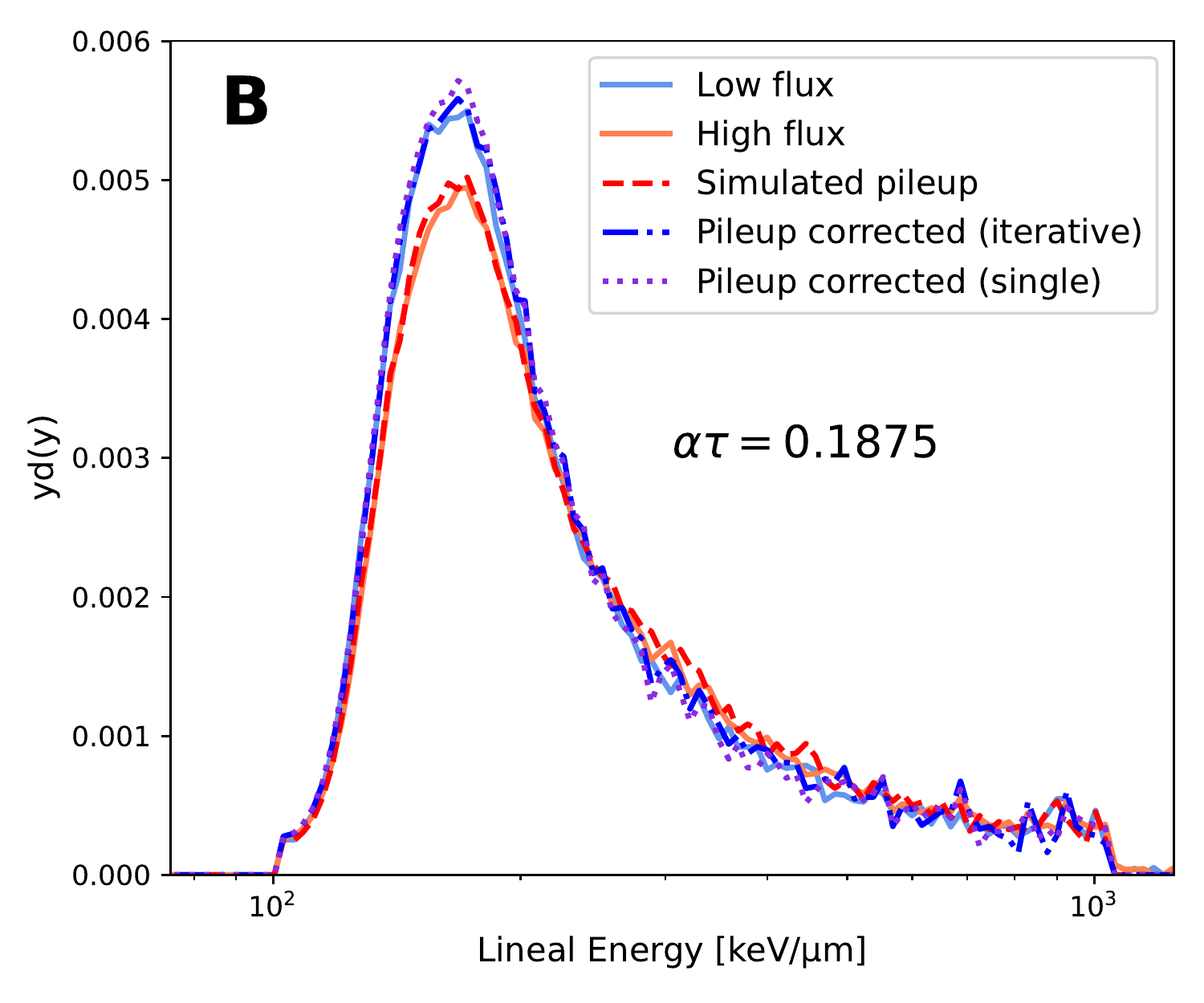}
\caption{ Measured high flux, low flux and pileup-corrected spectra for \ce{^12C^6+} ions at 120 MeV/u. Both the pileup simulations and the corrected spectra are depicted with dashed lines. The spectra were measured without online pileup rejection (PUR). (A) Spectrum measured in the plateau region. (B) Spectrum measured at the Bragg peak. \label{fig:results_120MeV}}
\end{figure}

\begin{figure}[htbp]
\centering
\includegraphics[width=.45\textwidth]{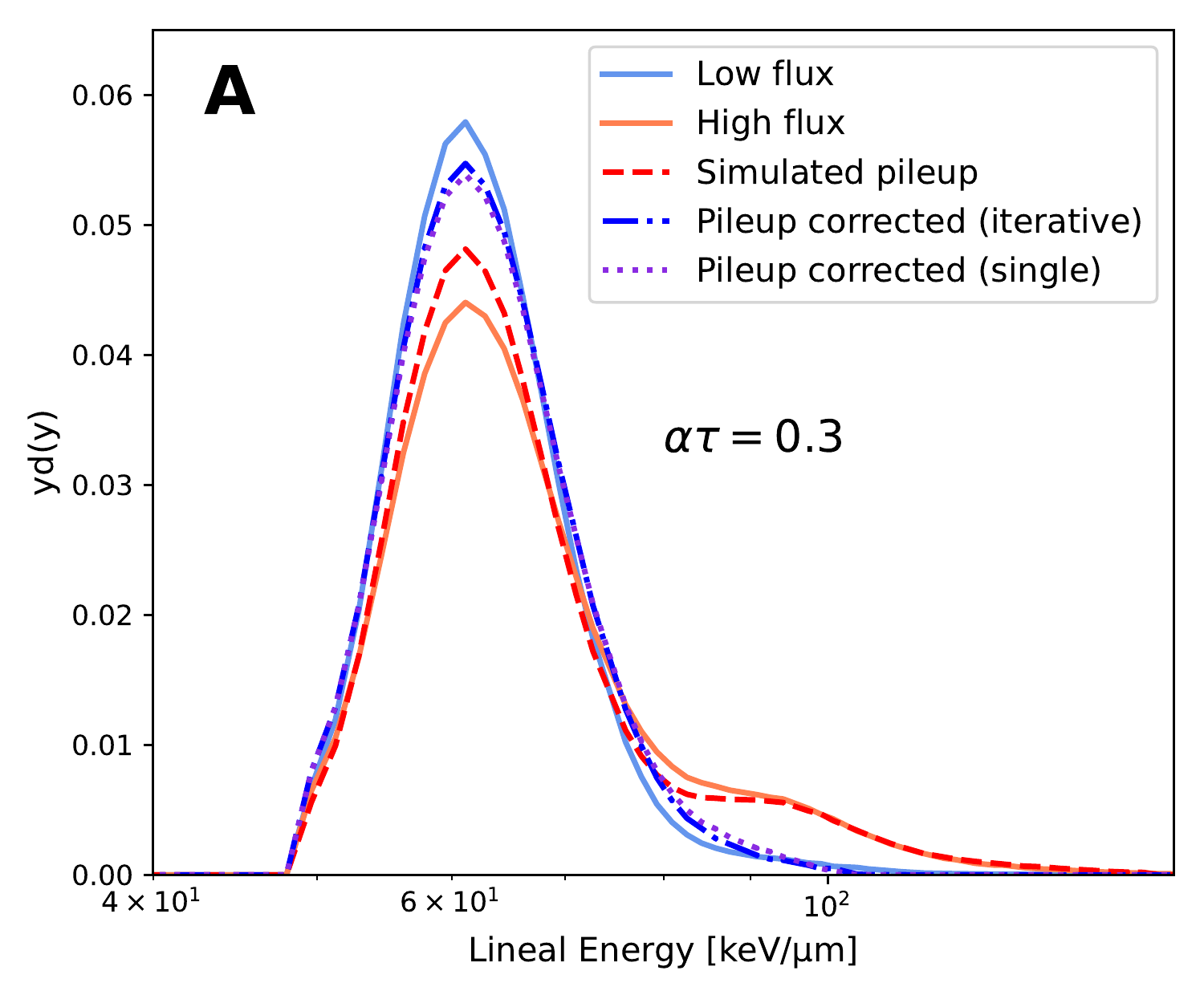}
\qquad
\includegraphics[width=.45\textwidth]{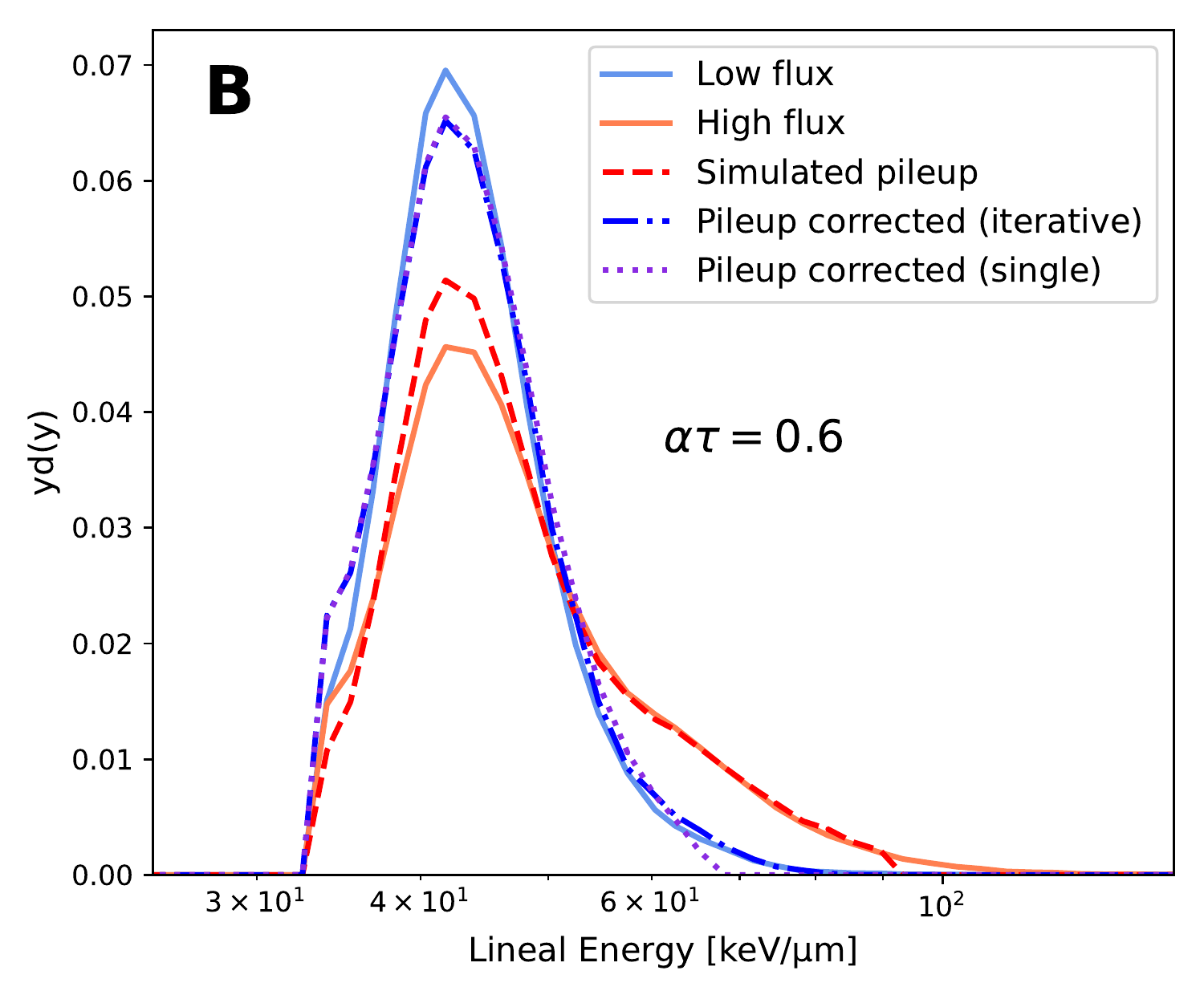}
\caption{Measured high flux, low flux and pileup-corrected spectra for \ce{^12C^6+} ions. Both the pileup simulations and the corrected spectra are depicted with dashed lines. The spectra were measured without online pileup rejection. (A) Spectrum measured in the plateau region at 120 MeV/u. (B) Spectrum measured in the plateau region at 238.6 MeV/u. \label{fig:results_first_shift}}
\end{figure}

Pileup is observed in all spectra measured at clinical rates (high flux, $\sim 4.3\cdot10^7$ particles/s). The results are shown in figure \ref{fig:results_120MeV} and \ref{fig:results_first_shift}. In the low flux measurements ($\sim 4.3\cdot10^6$ particles/s), pileup did not significantly alter the shapes of the measured pulses from the expected Landau/Gaussian distribution. This was previously confirmed using the same detector to measure at a very low flux carbon setting (238.6 MeV/u) with a particle rate on the order of \SI{}{\kilo\hertz}. The difference in the bin occupations between a normalized low flux and a normalized very low flux spectrum was quantified as $\sum \left ( \Delta k_i \right )^2\approx0.02$. The results are shown in figure ~\ref{fig:linear_axis_spec}A. The very low flux setting was not used for the measurements in this paper. Although not entirely pileup-free, the low flux spectra are considered the optimal endpoint for the pileup correction, as the relative difference between the low and extremely low flux spectra is negligible.\\

The pulse shapes in figure ~\ref{fig:linear_axis_spec}B illustrate the difference in pileup between the high and low flux spectra measured for the analysis. A double sum peak is emerging at twice the most probable value (MPV) channel 110 in the high flux spectra, indicating pulse pileup. During the presented measurements, both spectra were obtained with and without the internal pileup rejection logic (PUR) of the shaping amplifier. As shown in figure ~\ref{fig:linear_axis_spec}, the rejection circuit was not capable of prohibiting the acquisition of pileup events. The analysis was therefore performed without PUR and figure \ref{fig:results_120MeV} and \ref{fig:results_first_shift} show versions of the spectra without internal rejection.\\

As shown in figure \ref{fig:results_120MeV} and \ref{fig:results_first_shift}, the resampling algorithm is able to realistically recreate the pileup encountered in the measured high flux spectra from the measured low flux spectra. Moreover, it is able to effectively correct the high flux spectra, so they conform with the measured low flux spectra. For a fixed readout setup and particle energy, it was possible to simulate and correct pileup in two spectra measured at different positions along the depth-dose curve using the same parameter $\alpha \tau$. For a different amplifier gain and particle energy, new $\alpha \tau$ needed to be established.\\

Figure \ref{fig:results_120MeV} shows the pulse height spectra measured in the plateau region and at the Bragg peak (at a depth of \SI{28.3}{\milli\meter}) of a monoenergetic 120 MeV/u \ce{^12C^6+} ion beam, which were both successfully corrected to resemble the low flux measurement. The correction was possible in a single step with $\alpha \tau = 0.1875$. An iterative correction using $\alpha\tau= 0.0375$ converged after 5 steps in both cases. This is evident by the double sum pulse in the tail region disappearing. Due to the wider energy range covered in the spectrum measured near the Bragg peak, the visual effect of pileup in the tail is not as pronounced as it is in the plateau spectra.\\

In a previous measurement campaign, the same measurement setup was used with the shaping amplifier set to a different gain, measuring spectra in the plateau region using both 120 MeV/u and 238.6 MeV/u \ce{^12C^6+} ions. Different pileup magnitudes were used to simulate and correct pileup for these measurements. Specifically, $\alpha\tau=0.3$ was established for the 120 MeV/u and $\alpha\tau=0.6$ for the 238.6 MeV/u measurement. The iterative correction converged in 8 steps for the 120 MeV/u measurement and in 14 steps for 238.6 MeV/u using $\alpha\tau= 0.0375$. Both measured and corrected spectra are shown in figure \ref{fig:results_first_shift}.\\

\section{Discussion}

To compare the spectra, the frequency average lineal energy $\bar{y}_F$ and the dose average lineal energy $\bar{y}_D$ values are taken as indicators for a global change in the data. This change is quantified as the relative difference in these values. The comparison is shown in table \ref{tab:means}. The measured high flux spectra are taken as the reference for the pileup simulation, and the measured low flux spectra are taken as a reference for the pileup corrected spectra. Without any correction, the average relative differences in the mean values between the high and low flux spectra in all four measurements presented in figure \ref{fig:results_120MeV} and figure \ref{fig:results_first_shift} are $\Delta_\% \bar{y}_F$=\SI[separate-uncertainty=true]{3.57(1.92)}{\percent} and $\Delta_\% \bar{y}_D$=\SI[separate-uncertainty=true]{5.80(2.73)}{\percent}. Since the only variable that changed between measurements was the particle rate, any observed change can be solely attributed to pileup. Using the single step correction, using the predetermined $\alpha\tau$ obtained from the forward simulation, this can be reduced to $\Delta_\% \bar{y}_F$=\SI[separate-uncertainty=true]{0.61(0.28)}{\percent} and $\Delta_\% \bar{y}_D$=\SI[separate-uncertainty=true]{1.31(0.94)}{\percent}. With the iterative correction it was possible to reach $\Delta_\% \bar{y}_F$=\SI[separate-uncertainty=true]{0.28(0.08)}{\percent} and $\Delta_\% \bar{y}_D$=\SI[separate-uncertainty=true]{0.97(0.66)}{\percent}. \\

\begin{table}
\caption{Calculated microdosimetric mean values $\bar{y}_F$ and $\bar{y}_D$ of the measured and corrected spectra. The relative differences to the corresponding high and low flux measurements are indicated. Both for the simulation of pileup into the low flux spectra and the correction of pileup from the high flux spectra, a minimal difference to the respective measurements could be achieved.}
\label{tab:means}

\resizebox{\textwidth}{!}{
\begin{tabular}{l|lcccc}
\br

\textbf{Setup} &
  \textbf{Type} &
  \textbf{$\bar{y}_F$ {(}keV/µm{)}} &
  \textbf{$\bar{y}_D$ {(}keV/µm{)}} &
  \textbf{\begin{tabular}[c]{@{}l@{}}Rel. Difference to\\ Measured $\bar{y}_F$ {(}\%{)}\end{tabular}} &
  \textbf{\begin{tabular}[c]{@{}l@{}}Rel. Difference to\\ Measured $\bar{y}_D$ {(}\%{)}\end{tabular}} \\ \hline
  
\multirow{5}{*}{\begin{tabular}[c]{@{}l@{}}\ce{^12C^6+} 120 MeV/u\\Plateau (Figure \ref{fig:results_120MeV})\end{tabular}} & High flux                & 63.02  & 65.24  & -    & -    \\
                                                                                                   & Simulated Pileup        & 63.01  & 64.66  & 0.02 & 0.89 \\ \cline{2-6}
                                                                                                   & Low flux                 & 61.49  & 62.98  & -    & -    \\ 
                                                                                                   & Corrected (iterative)   & 61.6  & 63.26  & 0.18 & 0.44 \\
                                                                                                   & Corrected (single step) & 61.32  & 62.32   & 0.28 & 1.05 \\ \hline
\multirow{5}{*}{\begin{tabular}[c]{@{}l@{}}\ce{^12C^6+} 120 MeV/u\\Peak (Figure \ref{fig:results_120MeV})\end{tabular}}       & High flux                & 203.52 & 258.17 & -    & -    \\
                                                                                                   & Simulated Pileup        & 203.30 & 256.31 & 0.11 & 0.72 \\ \cline{2-6}
                                                                                                   & Low flux                 & 197.46 & 248.21 & -    & -    \\
                                                                                                   & Corrected (iterative)   & 196.76 & 244.02 & 0.35 & 1.69 \\
                                                                                                   & Corrected (single step) & 195.80 & 243.67 & 0.84 & 1.83 \\ \hline
\multirow{5}{*}{\begin{tabular}[c]{@{}l@{}}\ce{^12C^6+} 120 MeV/u\\Plateau (Figure \ref{fig:results_first_shift})\end{tabular}}    & High flux                & 65.09  & 67.87  & -    & -    \\
                                                                                                   & Simulated Pileup        & 64.62  & 66.75  & 0.72 & 1.65 \\ \cline{2-6}
                                                                                                   & Low flux                 & 62.12  & 63.44  & -    & -    \\
                                                                                                   & Corrected (iterative)   & 62.28  & 63.20  & 0.26 & 0.38 \\
                                                                                                   & Corrected (single step) & 62.42  & 63.37  & 0.48 & 0.11 \\ \hline
\multirow{5}{*}{\begin{tabular}[c]{@{}l@{}}\ce{^12C^6+} 238.6 MeV/u\\Plateau (Figure \ref{fig:results_first_shift})\end{tabular}}  & High flux                & 46.31  & 49.31  & -    & -    \\
                                                                                                   & Simulated Pileup        & 46.04  & 48.03  & 0.58 & 2.60 \\ \cline{2-6}
                                                                                                   & Low flux                 & 43.20  & 44.70  & -    & -    \\
                                                                                                   & Corrected (iterative)   & 43.06  & 44.08  & 0.32 & 1.39 \\
                                                                                                   & Corrected (single step) & 42.83  & 43.69  & 0.86 & 2.26

\end{tabular}
}
\end{table}

Nevertheless, limitations have to be acknowledged. The effective rate parameter $\alpha$ is highly dependent on the beam optics, the acquisition setup and the radiation quality, which is why it needs to be empirically determined for different measurement settings. A precise value cannot be achieved by a simplistic Poissonian approach, as the assumption of a constant average arrival of particle does not hold true for accelerator beams, as recently discussed by ~\cite{Torino_Rate_Meas}. While a preliminary estimate of $\alpha$ can be made through the average particle rate according to  \eqref{eq:ion_beam_est}, this value does not necessarily represent the parameter required for correction. For the high flux measurements using the 120 MeV/u carbon ions, the estimation in equation \eqref{eq:ion_beam_est} gives an effective average rate of $1.68\cdot10^4$ particles/s crossing the detector surface. For 120 MeV/u this would result in a pileup magnitude of $\alpha\tau=0.126$. Using $\alpha\tau=0.126$, however, neither adequately simulates pileup in the low flux spectra nor effectively corrects pileup in the high flux spectra. Given these complexities, a uniform correction factor cannot be applied across different measurement settings. The estimation for 238.6 MeV/u carbon ions gives $3.38\cdot10^4$ particles/s, due to the smaller beam spot ~\cite{MAUS_Carbon_Comissioning}. For 238.6 MeV/u, the expected pileup magnitude is $\alpha\tau=0.254$, also deviating from the pileup magnitude needed for an effective correction, $\alpha\tau=0.6$.\\

Nevertheless, the empirical determination of the value ensures that the correction parameters are tailored to the specific experimental conditions. Stochastic resampling has demonstrated the capacity to improve spectra and provides a practical means for offline pileup correction using a traditional analog measurement setup. In regularly repeated measurements, as for example in beam quality assurance ~\cite{Bolst_2018_QA}, the offline correction method can be used to optimize the balance between measurement time, statistical accuracy, and minimizing pileup. By establishing the correction parameters once using a high-statistics, pileup-free spectrum, subsequent high flux measurements showing moderate pileup can be measured and corrected at short timescales. For future online rejection systems, different strategies including digital pulse processing are being investigated, depending on an increased \gls{SNR} and high throughput. In order to reduce pileup to an insignificant level, it is necessary to construct a readout system that is specifically designed to meet the unique requirements of microdosimetric measurements at high dose rates.

\section{Conclusion}
\label{sec:conclusion}

Measuring precise energy deposition spectra at high particle rates requires an effective rejection or correction of pileup. The conventional approach of limiting the detector area and employing traditional pileup-rejection systems, such as fast-slow coincidence, is insufficient to avoid pileup in microdosimetric measurements using solid-state microdosimeters at clinical dose rates. While online rejection would be ideal, current hardware limitations necessitate alternative solutions. Post-processing methods present a promising tool as they do not require specialized readout equipment. A resampling algorithm ~\cite{Langen} was adapted and the capability to correct pileup was demonstrated in four separate measurements. The relative change in the spectra due to pileup distortions was evaluated using the frequency mean $\bar{y}_F$ and dose mean $\bar{y}_D$ values. Using iterative correction, the initial average relative differences $\Delta_\% \bar{y}_F$=\SI[separate-uncertainty=true]{3.57(1.92)}{\percent} and $\Delta_\% \bar{y}_D$=\SI[separate-uncertainty=true]{5.80(2.73)}{\percent} due to pulse pileup could be reduced to $\Delta_\% \bar{y}_F$=\SI[separate-uncertainty=true]{0.28(0.08)}{\percent} and $\Delta_\% \bar{y}_D$=\SI[separate-uncertainty=true]{0.97(0.66)}{\percent}. The main limitation of the method presented is the need to empirically determine the correction parameters for each setup. The amount of correction is controlled via the pileup magnitude $\alpha \tau$. While $\tau$ only depends on the unit pulse shape of the shaping network, $\alpha$ is a complex parameter incorporating information about the particle rate, beam characteristics and measurement choices, which makes it difficult to estimate \textit{a priori}.\\

Offline correction could be a valuable tool for quality assurance measurements, since measurements are performed under similar conditions and the correction parameters can be established in advance. The long-term objective should be the development of readout systems capable of rejecting pileup at clinical dose rates. In situations where pulse pileup cannot be avoided, an offline resampling algorithm, as presented in this work, can be an effective tool for spectrum correction.

\ack
The financial support of the Austrian Ministry of Education, Science and Research is gratefully acknowledged for providing beam time and research infrastructure at MedAustron. The authors would also like to express their sincere gratitude to Dr. Claudio Verona and the Department of Industrial Engineering of Rome, Tor Vergata University for their instrumental support in providing the diamond detector used for the measurements presented in this manuscript.

\section*{References}
\begin{harvard}
\item \bibliography{bib}
\end{harvard}

\end{document}